\documentclass[12pt]{article}
\usepackage[utf8]{inputenc}
\usepackage[top=50pt,bottom=50pt,left=68pt,right=66pt]{geometry}
\usepackage{amsmath}
\usepackage{booktabs}
\usepackage{amssymb}
\usepackage[title]{appendix}
\usepackage{mathrsfs}
\usepackage{cite}
\usepackage{float}
\usepackage{xcolor}
\usepackage{multirow}
\usepackage{graphicx,caption,subcaption}
\usepackage[space]{grffile}
\newcommand{\overbar}[1]{\mkern1.5mu\overline{\mkern-1.5mu#1\mkern-1.5mu}\mkern 1.5mu}

\definecolor{darkblue}{rgb}{0.1,0.1,.7}
\usepackage[breaklinks=true,linktocpage=true,
colorlinks=true,urlcolor=darkblue,linkcolor=darkblue,
citecolor=darkblue,pdfpagelabels=true,hypertexnames=true,
plainpages=false,naturalnames=false,]{hyperref}

\def\co{{\cal O}}

\def\car{{\cal R}}

\def\bs{{\bf S}}

\begin{document}

\renewcommand{\arraystretch}{1.3}
\thispagestyle{empty}

\noindent {\hbox to\hsize{
\vbox{\noindent March 2023 \hfill IPMU22-0066}}
\noindent  $~{}$ revised version \hfill }


\noindent
\vskip2.0cm
\begin{center}

{\Large\bf Primordial black holes from Volkov--Akulov--Starobinsky\\
\vglue.1in
supergravity}

\vglue.3in

\hspace{-2pt}Yermek Aldabergenov~${}^{a,b,}$\footnote{yermek.a@chula.ac.th} and Sergei V. Ketov~${}^{c,d,e,}$\footnote{ketov@tmu.ac.jp}
\vglue.1in

${}^a$~{\it Department of Physics, Faculty of Science, Chulalongkorn University,\\ Phayathai Road, Pathumwan, Bangkok 10330, Thailand}\\
${}^b$~{\it Department of Theoretical and Nuclear Physics, 
Al-Farabi Kazakh National University,\\ 71 Al-Farabi Ave., Almaty 050040, Kazakhstan}\\
${}^c$~{\it Department of Physics, Tokyo Metropolitan University\\
1-1 Minami-ohsawa, Hachioji-shi, Tokyo 192-0397, Japan}\\
${}^d$~{\it Research School of High-Energy Physics, Tomsk Polytechnic University\\
2a Lenin Avenue, Tomsk 634028, Russian Federation}\\
${}^e$~{\it Kavli Institute for the Physics and Mathematics of the Universe (WPI)
\\The University of Tokyo Institutes for Advanced Study, Kashiwa 277-8583, Japan}\\
\vglue.1in

\end{center}

\vglue.3in

\begin{center}
{\Large\bf Abstract}
\vglue.2in
\end{center}

We study the formation of primordial black holes (PBH) in the Starobinsky supergravity coupled to the nilpotent superfield describing Volkov--Akulov goldstino. By using the no-scale K\"ahler potential and a polynomial superpotential, we find that under certain conditions our model can describe effectively single-field inflation with the ultra-slow-roll phase that appears near a critical  (near-inflection) point of the scalar potential. This can lead to the formation of PBH as part of (or whole) dark matter, while keeping the inflationary spectral tilt and the tensor-to-scalar ratio in good agreement with the current cosmic microwave background (CMB) bounds. After inflation, supersymmetry is spontaneously broken at the inflationary scale with the vanishing cosmological constant.

\newpage

\tableofcontents

\setcounter{footnote}{0}

\section{Introduction}

The nilpotent ($N=1$) superfields can be used to describe the low-energy effective field theories with spontaneously broken and non-linearly realized supersymmetry (SUSY)\cite{Rocek:1978nb,Ivanov:1978mx,Lindstrom:1979kq,Ivanov:1982bpa,Samuel:1982uh,Casalbuoni:1988xh,Hatanaka:2003cr,Komargodski:2009rz} in four spacetime dimensions. The nilpotency condition $\bs^2=0$ on a chiral superfield
\begin{equation}
\bs=S+\sqrt{2}\theta\chi+\theta^2F^S
\end{equation}
has a solution
\begin{equation}\label{sol}
S=\frac{\chi^2}{2F^S}~,
\end{equation}
where $S$ is the complex scalar, $\chi$ is the chiral fermion, and $F^S$ is the complex auxiliary field. This solution is consistent only if $F^S\neq 0$. The $\chi$ can be identified with {\it goldstino} of the broken $N=1$ SUSY. As regards inflationary dynamics, only scalar fields are relevant, hence, we can ignore the terms proportional to $S$ or $\chi^2$ in the scalar potential. The theory of a single nilpotent chiral superfield $\bs$ is known to be equivalent \cite{Rocek:1978nb,Lindstrom:1979kq,Hatanaka:2003cr,Kuzenko:2010ef} to the Volkov-Akulov (VA) theory \cite{Volkov:1973ix}.

Antoniadis, Dudas, Ferrara and Sagnotti proposed a Starobinsky-like model of inflation with the no-scale K\"ahler potential, coupled to the VA theory by using the nilpotent chiral superfield, which was dubbed the Volkov-Akulov-Starobinsky (VAS) supergravity \cite{Antoniadis:2014oya}. During inflation, the VAS model \cite{Antoniadis:2014oya} is consistent with the nilpotency constraint because $F^S\neq 0$. However, the $F^S$ vanishes in a Minkowski vacuum, which makes the solution $S=\chi^2/(2F^S)$ singular. The improved and generalized version of the VAS model with a non-vanishing $F^S$ in vacuum also, was proposed in Ref.~\cite{Aldabergenov:2020atd} with the same no-scale K\"ahler potential. The alternative VAS model with consistent vacuum structure but a different K\"ahler potential was studied in Ref.~\cite{DallAgata:2014qsj}. 

On the other hand, primordial black holes (PBH) is an interesting area of research that can give us more information about the inflationary and post-inflationary epoch. The absence of PBH signals in cosmological and astrophysical observations can constrain our models, whereas, if found, PBH masses and their distribution could teach us important details about the mechanisms of PBH production. There is also an intriguing possibility that PBH of certain masses make up the observed dark matter (DM), see e.g.,
Refs.~\cite{Domenech:2021ztg,Escriva:2022duf} for a review.

In this paper we consider PBH (as part of or whole DM) from the VAS supergravity.~\footnote{As regards PBH formation in other Starobinsky-like supergravity-based inflationary models like the $\alpha$-attractors, see e.g., Refs.~\cite{Dalianis:2018frf,Mahbub:2019uhl,Aldabergenov:2020bpt,Nanopoulos:2020nnh,Ketov:2021fww,Aldabergenov:2022rfc,Frolovsky:2022qpg,Heurtier:2022rhf,Braglia:2022phb}.} We generalize the model of Ref.~\cite{Aldabergenov:2020atd} by extending the polynomial superpotential in order to derive the necessary conditions for the PBH production after inflation, and estimate PBH masses and abundance for the present DM. We employ the standard scenario of PBH formation from single-field models of inflation,
based on engineering a near-inflection point in the inflaton scalar potential leading to the ultra-slow-roll phase of inflation and the enhancement (peak) in the power spectrum of scalar curvature perturbations \cite{Garcia-Bellido:2017mdw,Germani:2017bcs}.
 Then the emerging large perturbations gravitationally collapse into PBH. We do not address non-Gaussianities and loop corrections in this paper, see Refs.~\cite{DeLuca:2022rfz,Riotto:2023hoz,Choudhury:2023vuj} for their possible impact.

Our paper is organized as follows. Sec.~2 is our setup, where we recover the original Starobinsky model, and demonstrate how to create a stationary near-inflection point for the ultra-slow-roll (USR) phase. In Sec.~3 we numerically derive  inflationary solutions and show viability of our new model. The main part is Sec.~4 devoted to PBH formation from enhanced scalar perturbations and the related PBH-DM scenario. We also find the scalar-induced gravitational waves (GW) spectrum that can be tested by future space-based GW experiments. In Appendix we briefly summarize the technical details related to Mukhanov-Sasaki (MS) equation. We set the reduced Planck mass $M_{\rm Pl}=1$ unless it is stated otherwise.

\section{Setup}\label{sec_setup}

The original VAS supergravity model \cite{Antoniadis:2014oya} uses the no-scale K\"ahler potential for the inflaton chiral superfield $T$ and the nilpotent superfield $S$ as
\begin{equation}\label{K_no-scale}
K=-3\log(T+\overbar T-S\overbar S)~,
\end{equation} 
with the superpotential 
\begin{equation}
W=S(b_0+b_1 T)~,
\end{equation}
having constant parameters $b_0$ and $b_1$. Since $S^2=0$, both $K$ and $W$ are linear with respect to $S$ and ${\bar S}$. 
Though this model does describe SUSY breaking during inflation, SUSY is restored in a Minkowski minimum, leading to $F^S=0$ that is inconsistent with the solution (\ref{sol}). The VAS model \cite{Antoniadis:2014oya} was improved in 
Ref.~\cite{Aldabergenov:2020atd} by extending the superpotential to
\begin{equation}\label{W_old}
W=A[a_0+a_1 T+S(b_0+b_1 T)]
\end{equation}
under certain conditions on the parameters to get a non-vanishing vacuum expectation value (VEV) $\langle F^S\rangle$. The parameter $A$ can be used to rescale one of the non-vanishing parameters  $a$ or $b$ to unity. All the parameters of the superpotential  (\ref{W_old}) are chosen to be real.

Our aim is to study further extensions of the superpotential (\ref{W_old}) toward PBH production in the VAS framework by adding an ultra-slow-roll (USR) regime in the effective single-field inflation scenario that requires the scalar potential to have a critical (near-inflection) point \cite{Garcia-Bellido:2017mdw, Motohashi:2017kbs}. The K\"ahler potential \eqref{K_no-scale} will be 
unchanged.

Let us consider a more general superpotential 
\begin{equation}\label{W_gen}
W=A[f(T)+Sg(T)]~,
\end{equation}
where $f(T)$ and $g(T)$ are polynomials in $T$. The F-type scalar potential is given by
\begin{equation}
V_F=e^K\left[K^{I\bar J}D_IWD_{\bar J}\overbar W-3W\overbar W\right]~,
\end{equation}
where $K^{I\bar J}$ is the inverse K\"ahler metric, the indices $I,J$ run over the chiral superfields, and $D_IW=W_I+K_IW$, with
the subscripts denoting the derivatives. The auxiliary $F$-fields are given by
\begin{equation}
F^I=-e^{K/2}K^{I\bar J}D_{\bar J}\overbar W~.
\end{equation}

In our model with the single chiral nilpotent superfield $S$, the K\"ahler metric becomes diagonal after using the constraint $S^2=0$ and ignoring the fermions, with $K_S=0$. This implies that $F^S\propto \overbar W_{\bar S}\neq 0$ for the consistency of the solution with the constraint. We also introduce the K\"ahler-invariant $F$-fields as
\begin{equation}
|F^S|\equiv\sqrt{K_{S\bar S}F^S\overbar F^S}~,~~~|F^T|\equiv\sqrt{K_{T\bar T}F^T\overbar F^T}~.
\end{equation}

\subsection{Starobinsky-like inflation}

There are several ways to realize Starobinsky inflation \cite{Starobinsky:1980te} in supergravity, as well as in our models, see e.g., Ref.~\cite{Ketov:2021fww} for a review and Ref.~\cite{Ivanov:2021chn} for possible extensions. Here we use the no-scale K\"ahler potential \eqref{K_no-scale} and the superpotential
\begin{equation}\label{W_star}
W=A[a_0+a_1T+a_2T^2+a_3T^3+S(b_0+b_1T+b_2T^2)]~,
\end{equation}
where we have expanded the functions $f(T)$ and $g(T)$ of Eq.~(\ref{W_gen}) in Taylor series up to the cubic and quadratic terms, respectively. Equation (\ref{W_star})  leads to the scalar potential $V=V_F$ given by
\begin{align}\label{V_setup}
\begin{aligned}
\frac{12}{A^2}V &=(b_0^2-6a_0a_1)\phi^{-2}+2(b_0b_1-2a_1^2-6a_0a_2)\phi^{-1}\\
&+(b_1^2+2b_0b_2-10a_1a_2-18a_0a_3)+2(b_1b_2-2a_2^2-6a_1a_3)\phi+(b_2^2-6a_2a_3)\phi^2~,
\end{aligned}
\end{align}
where we have used the parametrization
\begin{equation}
T=\phi+i\tau~,
\end{equation}
and have set the axion/sinflaton $\tau=0$ by assuming it to be stabilized, as will be shown below. 

The canonical parametrization of the inflaton is given by~\footnote{The sign in front of $\varphi$ is arbitrary, and we choose the negative sign.}
\begin{equation}\label{canon}
\phi=\langle\phi\rangle\exp\left( -\sqrt{2/3}\varphi \right)~,
\end{equation}
so that $\phi$ is always positive, while in vacuum we always have $\varphi=0$. 

When looking at Eq.~\eqref{V_setup}, one finds several ways to obtain a Starobinsky-like plateau, i.e. a nearly-flat inflaton potential. One option is to keep only negative powers of $\phi$ (and a constant term) by eliminating the positive powers, which can be done by the appropriate choice of the parameters. In this case the potential asymptotically approaches a constant value at $\varphi\rightarrow -\infty$. Another option is to keep only positive powers of $\phi$, so that the potential approaches a constant at $\varphi\rightarrow +\infty$. It is also possible to keep both positive and negative powers of $\phi$, and choose the parameters in Eq.~(\ref{V_setup}) for viable inflation.~\footnote{An example of such potential, inspired by string theory, was proposed in Ref.~\cite{Cicoli:2018asa}.} In the latter case, the potential becomes infinite when $\varphi\rightarrow \pm\infty$. We consider the first two options to the end of this Subsection, and then (in the next Subsection) activate the other parameters in Eq.~(\ref{V_setup}) for the purpose of PBH production and agreement with precision measurements of the tilt $n_s$ of (CMB) scalar perturbations, by allowing both positive and negative powers of $\phi$.

{\bf Inflation for negative $\varphi$.} Starobinsky-like inflation for large negative $\varphi$ (i.e. large values of $\phi$) can be obtained from the potential \eqref{V_setup} by arranging the coefficients at $\phi$ and $\phi^2$ to be zero. For example, this can be done by setting $a_2=b_2=0$, and either $a_1=0$ or $a_3=0$ (in Ref.~\cite{Aldabergenov:2020atd}, only the case of $a_2=b_2=a_3=0$ was considered). The higher-order terms ($a_4,a_5,b_3,b_4,$ etc.) are prohibited because they would lead to positive powers of $\phi$ in the scalar potential, which can unflatten the inflationary plateau unless the corresponding parameters are extremely small. These restrictions on the parameters make it difficult to create an inflection point in the potential in order to realize USR inflation and PBH production. Hence, we consider another scenario with inflation taking place for large positive $\varphi$ or small $\phi$.~\footnote{In our models, inflation is always of the single-large-field-type in terms of the canonical inflaton $\varphi$.}

{\bf Inflation for positive $\varphi$.} In this case we keep only positive powers of $\phi$ in Eq.~\eqref{V_setup}, so that the potential approaches a constant value at $\varphi\rightarrow +\infty$, which corresponds to $\phi\rightarrow 0$. This can be done by setting $a_1=b_0=0$, and either (I) $a_0=0$ or (II) $a_2=0$ (or both). In the first case, $a_1=b_0=a_0=0$, the scalar potential reads 
\begin{equation}\label{V_I}
\frac{12}{A^2}V=b_1^2+2(b_1b_2-2a_2^2)\phi+(b_2^2-6a_2a_3)\phi^2~.
\end{equation}
We can further simplify the model by setting either $b_2=0$ (model I-a) or $a_3=0$ (model I-b). In the former case, the Minkowski vacuum equations $V=0$ and $\partial_\phi V\equiv V_\phi=0$ yield
\begin{equation}
a_3=-\frac{2a_2^3}{3b_1^2}~,~~~\langle\phi\rangle=\frac{b_1^2}{2a_2^2}~.
\end{equation}
Without loss of generality we can set $a_2=1$ (by rescaling $A$ and other parameters), which also leads to $a_3<0$. By using the canonical parametrization $\phi=\langle\phi\rangle e^{-\sqrt{2/3}\varphi}=b_1^2e^{-\sqrt{2/3}\varphi}/2$, we get the Starobinsky potential
\begin{equation}\label{V_star}
V=\frac{A^2b_1^2}{12}\left(1-e^{-\sqrt{\frac{2}{3}}\varphi}\right)^2~.
\end{equation}

As was already mentioned above, consistency of our construction requires $D_SW=W_S\neq 0$ over the whole inflationary history, which is satisfied both during and after inflation for the parameter choice I. In particular, in the Minkowski vacuum we get 
$\langle W_S\rangle=-Ab_1^3/2$, while the K\"ahler-invariant F-terms are given by
\begin{equation}
\langle |F^S|\rangle=\langle |F^T|\rangle=\frac{Ab_1}{2\sqrt{3}}~.
\end{equation}

We also find that the axion $\tau$ is stabilized with the vanishing VEV in the Minkowski vacuum and the positive mass squared, $m^2_\tau=A^2b_1^2/9$ (after canonical rescaling), that can be chosen beyond the Hubble scale during inflation. The effective axion mass at the horizon exit can be roughly estimated in the limit  $e^{-\sqrt{2/3}\varphi}\rightarrow 0$, which yields $m_{\tau,{\rm eff.}}\simeq m_\tau$, i.e. it is near  the axion mass in the vacuum.

Another route to the Starobinsky potential (model I) is given by  the case I-b, where $a_3=0$ and $b_2\neq 0$. The vacuum equations for the potential \eqref{V_I} imply
\begin{equation}
b_2=\frac{a_2^2}{b_1}~,~~~\langle\phi\rangle=\frac{b_1^2}{a_2^2}~,
\end{equation}
while we can set $a_2=1$ by rescaling the parameter $A$. In terms of the canonical inflaton $\phi=\langle\phi\rangle e^{-\sqrt{2/3}\varphi}$, we obtain the same Starobinsky potential \eqref{V_star}. In this case, the axion mass is unchanged, $m_\tau^2=A^2b_1^2/9$, being also approximately equal to the effective mass during early inflation. SUSY is broken by the F-field VEV, which are slightly different from the I-a case,
\begin{equation}
\langle |F^S|\rangle=\tfrac{1}{\sqrt{3}}Ab_1~,~~~\langle |F^T|\rangle=\sqrt{\tfrac{2}{3}} Ab_1~.
\end{equation}

The model II uses $a_1=b_0=a_2=0$. It leads to the scalar potential
\begin{equation}
\frac{12}{A^2}V=b_1^2-18a_0a_3+2b_1b_2\phi+b_2^2\phi^2~,
\end{equation}
whose stationary point equation is solved by $b_1+b_2\phi=0$. This is, however, problematic because the F-field of the nilpotent superfield,
\begin{equation}
F^S\propto\overbar W_{\bar S}=\overbar T(b_1+b_2\overbar T)=\phi(b_1+b_2\phi)~,
\end{equation}
vanishes at the stationary point (the minimum) when the axion $\tau$ vanishes, which is inconsistent with the solution to the nilpotency constraint. Therefore, model II is excluded from our discussion.

{\bf Elimination of irrelevant parameters.} Given the polynomial superpotential in the form \eqref{W_gen}, two of its {\it non-vanishing} parameters can be eliminated by reparametrisation. Let us rewrite the superpotential as
\begin{equation}\label{W_gen2}
W=A\sum_{i=m}a_iT^i+AS\sum_{j=n}b_jT^j~,
\end{equation}
where the integers $m$ and $n$ are positive, and the summation upper limits are arbitrary but greater than the lower limits. We assume that the lowest-order parameters $a_m$ and $b_n$ are non-vanishing. The coefficient $a_m$ (or any one of the non-vanishing $a_i$-coefficients) can always be set to unity by rescaling $A$ and the other parameters accordingly, while $b_n$ can be eliminated by the following redefinitions of the fields and the parameters:
\begin{gather}\label{b_reparam}
\begin{gathered}
T\rightarrow Tb_n^k~,~~~S\rightarrow Sb_n^{k/2}~,\\
a_i\rightarrow a_ib_n^{k(m-i)}~,~~~b_j\rightarrow b_jb_n^{k(m-j-1/2)}~,\\
A\rightarrow Ab_n^{-km}~,~~~k\equiv(m-n-\tfrac{1}{2})^{-1}~,
\end{gathered}
\end{gather} 
for $i>m$ and $j>n$. The superpotential \eqref{W_gen2} is invariant under the transformation \eqref{b_reparam}, while the K\"ahler potential $K=-3\log(T+\overbar T-S\overbar S)$ is shifted by an irrelevant constant,
\begin{equation}\label{K_b_reparam}
K\rightarrow K-3\log b_n^k~,
\end{equation}
which can be absorbed by rescaling $W$. Therefore, we can fix the lowest-order non-vanishing parameters as $a_m=b_n=1$.

\subsection{Engineering a critical (near-inflection) point}

To accommodate a near-inflection point, in general, the potential must be at least cubic in $\phi$. However, due to the non-trivial structure of the scalar potential in supergravity, the higher-order terms may be also needed. We find that it is sufficient to add a quartic term to the superpotential, either $W\supset Aa_4T^4$ or $W\supset Ab_3ST^3$. We demonstrate that the ultimate superpotential must have the form~\footnote{We choose $a_2=1$, while $a_0$ must be small against the other parameters of the order one.}
\begin{equation}
W=A[a_0+T^2+a_3T^3+S(T+b_2T^2)]+\Delta W~,
\end{equation} 
where $\Delta W$ should be either $Aa_4T^4$ or $Ab_3ST^3$ (we consider both cases below). The non-vanishing  $a_0$-term is needed to obtain the observed value of the spectral tilt $n_s$ (within $1\sigma$ CL).

\begin{figure}
\centering
  \includegraphics[width=.45\linewidth]{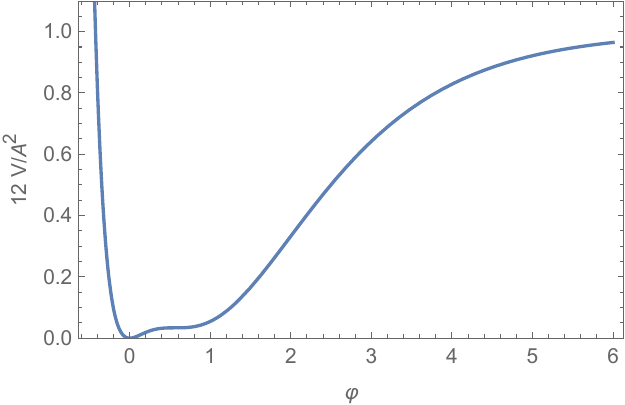}
\captionsetup{width=1\linewidth}
\caption{The scalar potential \eqref{V_IPM_minimal2} in the "minimal" model  with a near-inflection point. The inflection point is shallow and cannot be raised in this model.}
\label{Fig_V1}
\end{figure}

{\bf Model with $\Delta W=Aa_4T^4$.} In this case we begin with the minimal superpotential that allows an inflection point,
\begin{equation}\label{W_IPM_minimal}
W=A(T^2+a_3T^3+a_4T^4+ST)~,
\end{equation}
where we have $a_0=b_2=0$ at first, and further explain why $b_2\neq 0$ is needed. The impact of the parameter $a_0$ is  studied in Sec.~\ref{Sec_USR}. In Eq.~\eqref{W_IPM_minimal} we have two free parameters ($a_3$ and $a_4$) to create an inflection point. We also set a Minkowski minimum after inflation. The parameter $A$ is responsible for the height of the inflationary plateau and, therefore, the scale of inflation. The choice \eqref{W_IPM_minimal} leads to
\begin{equation}\label{V_IPM_minimal}
\frac{12}{A^2}V=1-4\phi-6a_3\phi^2-4a_4\phi^3+6a_3a_4\phi^4
+8a_4^2\phi^5~.
\end{equation}
The stationary point equation can be conveniently written as
\begin{equation}\label{V_phi_IPM_minimal}
\frac{3}{A^2}V_\phi=Z_1Z_2=0~;~~~Z_1\equiv -1+2a_4\phi^2~,~~~Z_1\equiv 1+3a_3\phi+5a_4\phi^2~,
\end{equation}
so that its four stationary points are the solutions to two quadratic equations $Z_1=0$ and $Z_2=0$. A stationary inflection point, which we call $\tilde\phi$, must satisfy the equations $V_\phi=V_{\phi\phi}=0$, which also fixes one of the parameters. Another parameter, along with $\langle\phi\rangle$, can be fixed by the Minkowski vacuum equations $V=V_\phi=0$. We want a 
near-inflection point to be between the horizon exit $\phi_*\approx 0$ and the vacuum $\langle\phi\rangle$, so we search for solutions to the two sets of equations satisfying this condition. We find the desired inflection point solves $Z_2=0$, and it is given by $\tilde\phi=-2/(3a_3)$ with the parameter $a_4=9a_3^2/20$ fixed by the equation $V_{\phi\phi}=0$. This implies that $a_3<0$ and $a_4>0$. Subsequently, using the Minkowski vacuum equations, we find
\begin{equation}
a_3=\frac{10-4\sqrt{10}}{3}~,~~~\langle\phi\rangle=-\frac{\sqrt{10}}{3a_3}~,
\end{equation}
where $\langle\phi\rangle$ solves $Z_1=0$. The scalar potential \eqref{V_IPM_minimal} with the canonical inflaton 
$\phi=\langle\phi\rangle e^{-\sqrt{2/3}\varphi}$ and the fixed $a_3$ and $a_4$ as above, reads
\begin{equation}\label{V_IPM_minimal2}
\frac{12}{A^2}V=\frac{2\sqrt{10}-5-2\sqrt{10}x+10x^2-\sqrt{10}x^3-5x^4+\sqrt{10}x^5}{2\sqrt{10}-5}~,~~~x\equiv e^{-\sqrt{\frac{2}{3}}\varphi}~.
\end{equation}
Its plot is shown in Fig.~\ref{Fig_V1}. The fact that there are no free parameters left in Eq.~\eqref{V_IPM_minimal2} means that we can tune the shape of the potential in the vicinity of the inflection point (by tuning $a_4$) but the height of the inflection point (against the slow-roll plateau) cannot be controlled. 

Our numerical analysis of the equations of motion shows that it is difficult to obtain an USR regime near the inflection point because of its shallow nature and the lack of control over the potential. However, the situation can be improved by turning on another parameter. We find that it is enough to turn on $b_2$. It extends the previous ``minimal" superpotential and the resulting scalar potential as follows:
\begin{gather}
W=A[T^2+a_3T^3+a_4T^4+S(T+b_2T^2)]~,\label{W_IPM_a4}\\
\frac{12}{A^2}V=1+2(b_2-2)\phi+(b_2^2-6a_3)\phi^2-4a_4\phi^3+6a_3a_4\phi^4
+8a_4^2\phi^5~.\label{V_IPM_a4}
\end{gather}

Given the non-vanishing $b_2$ parameter, the stationary equation $V_\phi=0$ loses its simple factorized form \eqref{V_phi_IPM_minimal} and becomes a more general quartic polynomial equation for its roots. However, we can get an approximate analytical solution by using our previous result, when $|b_2|\ll 1$. This is helpful to qualitatively study the behavior of the potential with  increasing $|b_2|$. For small $|b_2|$, we Taylor-expand $\tilde\phi$, $\langle\phi\rangle$, $a_4$, and $a_3$ as
\begin{align}
\begin{aligned}
\tilde\phi &=\tilde\phi_0+\tilde\phi_1b_2+\co(b_2^2)~,\\
\langle\phi\rangle &=\langle\phi\rangle_0+\langle\phi\rangle_1b_2+\co(b_2^2)~,\\
a_4 &=a_{4(0)}+a_{4(1)}b_2+\co(b_2^2)~,\\
a_3 &=a_{3(0)}+a_{3(1)}b_2+\co(b_2^2)~,
\end{aligned}
\end{align}
where $\tilde\phi_0$, $\langle\phi\rangle_0$, $a_{4(0)}$, and $a_{3(0)}$ are given by the ``unperturbed" ($b_2=0$) solutions for the potential \eqref{V_IPM_minimal},
\begin{equation}
\tilde\phi_0=-\frac{2}{3a_{3(0)}}~,~~~\langle\phi\rangle_0=-\frac{\sqrt{10}}{3a_{3(0)}}~,~~~a_{4(0)}=\frac{9a_{3(0)}^2}{20}~,~~~a_{3(0)}=\frac{10-4\sqrt{10}}{3}~.
\end{equation}
Solving the equations in the subleading order with respect to  $b_2$ yields
\begin{gather}
\begin{gathered}
\tilde\phi_1=\frac{18a_{3(1)}+5	a_{3(0)}}{27a_{3(0)}^2}~,~~~\langle\phi\rangle_1=\frac{16+\sqrt{10}}{216}~,\\
a_{4(1)}=\frac{3a_{3(0)}}{40}(12a_{3(1)}+5a_{3(0)})~,~~~a_{3(1)}=\frac{11\sqrt{10}-25}{9}~.
\end{gathered}
\end{gather}

\begin{figure}
\centering
  \includegraphics[width=.45\linewidth]{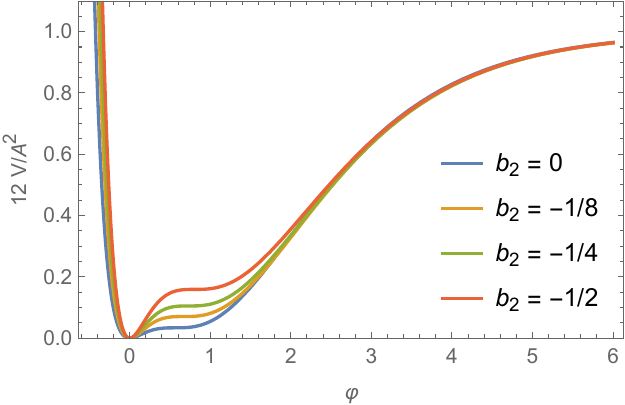}
\captionsetup{width=1\linewidth}
\caption{The scalar potential \eqref{V_IPM_a4} (in the model $a_4$) for some values of $b_2$. The $a_3$ and $a_4$ are derived from the inflection point and vacuum equations, while their values are collected in Table \ref{Tab_IPM_a4}.}
\label{Fig_V_IPM_a4}
\end{figure}

To get a dependence of the height of the inflection point upon small variations of $b_2$, we calculate the ratio $V(\tilde\phi)/V(0)$, i.e. the ratio of the value of $V$ at the inflection point to its asymptotic value at the slow-roll plateau where $\phi\simeq 0$. We find
\begin{equation}
\frac{V(\tilde\phi)}{V(0)}\approx 0.03-0.28b_2+\co(b_2^2)~.
\end{equation}
Hence, in order to raise the height of the inflection point, we need $b_2<0$. We confirm that by choosing  a small negative $b_2$ and numerically solving the inflection point (and the Minkowski vacuum) equations. This leads to the scalar potential shown in Fig.~\ref{Fig_V_IPM_a4}, where we set $b_1=1$ and choose different values of $b_2$. The corresponding parameter sets are shown in Table \ref{Tab_IPM_a4}.

\begin{table}[hbt!]
\centering
\begin{tabular}{l | r r r r}
\toprule
$b_2$ & $0$ & $-1/8$ & $-1/4$ & $-1/2$ \\
$a_3$ & $-0.8830$ & $-1.0271$ & $-1.1831$ & $-1.5166$ \\
$a_4$ & $0.3509$ & $0.4326$ & $0.5320$ & $0.7753$ \\
\bottomrule
\end{tabular}
\captionsetup{width=1\linewidth}
\caption{The parameter sets used in Fig.~\ref{Fig_V_IPM_a4} for the model with the scalar potential \eqref{V_IPM_a4}.}
\label{Tab_IPM_a4}
\end{table}

\begin{figure}
\centering
  \includegraphics[width=.45\linewidth]{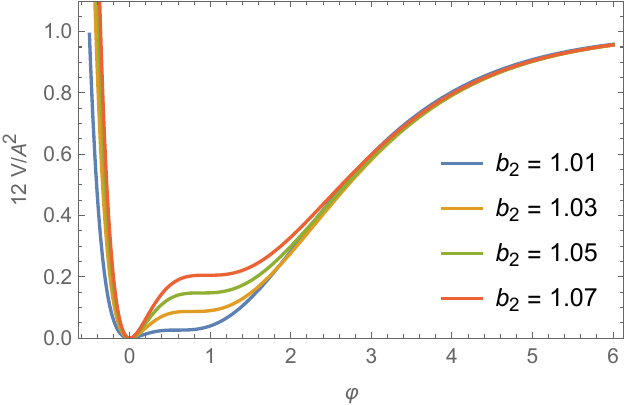}
\captionsetup{width=1\linewidth}
\caption{The scalar potential \eqref{V_IPM_b3} in the model $b_3$ with $b_2$ around unity. The values of $a_3$ and $b_3$ are shown in Table \ref{Tab_IPM_b3}. When $b_2\leq 1$, the inflection point vanishes.}
\label{Fig_V_IPM_b3}
\end{figure}

{\bf The model $\Delta W=Ab_3ST^3$.} In this model we have
\begin{equation}
W=A[T^2+a_3T^3+S(T+b_2T^2+b_3T^3)]~,
\label{W_IPM_b3}
\end{equation}
where we use the non-vanishing $b_2$-parameter, as is required by the inflection point and the Minkowski vacuum conditions. This leads to the following scalar potential:
\begin{equation}
\frac{12}{A^2}V=1+2(b_2-2)\phi+(b_2^2-6a_3+2b_3)\phi^2+2b_2b_3\phi^3+b_3^2\phi^4~.\label{V_IPM_b3}
\end{equation}
In this case, the inflection point and the vacuum are hard to find analytically, so we solve the equations numerically, by varying $b_2$ as a free parameter, and fixing $a_3$ and $b_3$ by the stationary points determined by the vacuum and inflection point equations. We find that the inflection point exists for $b_2$ slightly larger than one, as is shown in Fig.~\ref{Fig_V_IPM_b3}. The corresponding parameter sets are collected in Table \ref{Tab_IPM_b3}.

\begin{table}[hbt!]
\centering
\begin{tabular}{l | r r r r}
\toprule
$b_2$ & $1.01$ & $1.03$ & $1.05$ & $1.07$ \\
$a_3$ & $-0.1575$ & $-0.1559$ & $-0.1576$ & $-0.1606$ \\
$b_3$ & $-0.2399$ & $-0.2442$ & $-0.2533$ & $-0.2647$ \\
\bottomrule
\end{tabular}
\captionsetup{width=1\linewidth}
\caption{The parameter sets used in Fig.~\ref{Fig_V_IPM_b3} for the model with the scalar potential \eqref{V_IPM_b3}.}
\label{Tab_IPM_b3}
\end{table}

\section{Inflation and ultra-slow-roll}\label{Sec_USR}

In this Section we demonstrate viable inflation with a short USR period in our near-inflection-point models of supergravity. We begin with the model $a_4$ defined by Eqs.~\eqref{W_IPM_a4} and \eqref{V_IPM_a4}, and Fig.~\ref{Fig_V_IPM_a4}, where we fix $b_2=-1/2$. We denote the duration of the first (SR) and second (USR) inflationary stages as $\Delta N_1$ and $\Delta N_2$, respectively, and set the total inflation duration as $\Delta N_1+\Delta N_2=55$, by assuming the CMB reference scale $k=0.05~{\rm Mpc}^{-1}$ leaving the horizon $55$ e-folds before the end of inflation.

\begin{figure}
\centering
  \includegraphics[width=1\linewidth]{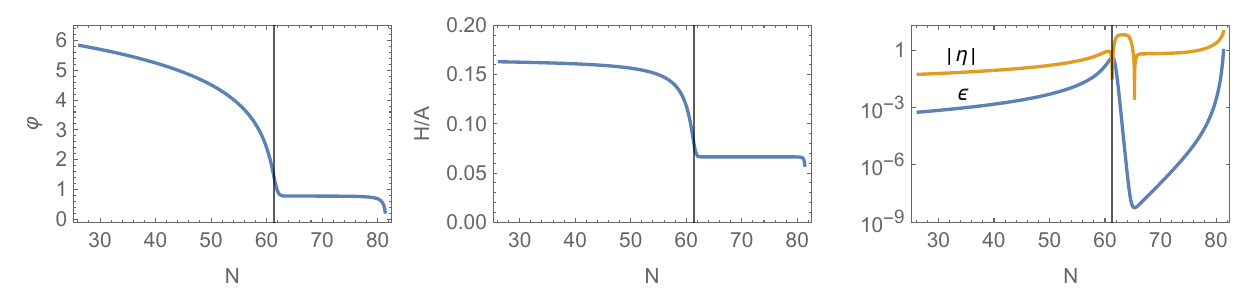}
\captionsetup{width=1\linewidth}
\caption{The numerical solutions to the equations of motion in terms of $\varphi(N)$ (on the left), $H(N)$ (in the center), and slow-roll parameters (on the right). The last $55$ e-folds are shown. The vertical line represents transition from SR to USR.  The duration of the USR stage is fixed to $\Delta N_2=20$.}
\label{Fig_a4_sol}
\end{figure}

For this purpose, we fix $\Delta N_2=20$ by adjusting the parameter $a_3$ around its inflection point value, while $a_4$ is to be fixed by the Minkowski vacuum equations. The desired outcome is obtained for
\begin{equation}
a_3=-1.5157647~,~~~a_4=0.774006~,\label{a3_a4_choice}
\end{equation}
and the corresponding inflationary solution is shown in Fig.~\ref{Fig_a4_sol}, which includes the inflaton evolution $\varphi(N)$, the Hubble function $H(N)$, and the Hubble slow-roll parameters 
\begin{equation}
\epsilon\equiv -\frac{H'}{H}~,~~~\eta\equiv\frac{\epsilon'}{\epsilon}~,
\end{equation}
during the last $55$ e-folds. The primes denote the derivatives with respect to $N$. The end of the first stage is defined by the local maximum of $\epsilon$ because it does not reach one at that time. The end of the second stage is defined by $\epsilon=1$. The initial conditions are set to $\varphi(0)=6.5$ and $\varphi'(0)=0.01$.

As is clear from Fig.~\ref{Fig_a4_sol}, the $\epsilon$ significantly dips during USR, as may be expected from the presence of a near-inflection point in the potential. This leads to a large enhancement in the scalar power spectrum. Before computing the power spectrum,  we derive the inflationary observables (cosmological tilts) $n_s$ and $r$ at the horizon exit (with $55$ e-folds before the end of inflation), by using the standard formulae
\begin{equation}
n_s=1-2\epsilon-\eta~,~~~r=16\epsilon~.
\end{equation}
For the model and the parameters under consideration, we find
\begin{equation}
n_s=0.9430~,~~~r=0.0093~.
\end{equation}
Therefore, the spectral tilt $n_s$ is outside the $3\sigma$ CMB limits \cite{BICEP:2021xfz,Tristram:2021tvh}. This problem can be solved by turning on one of the subleading parameters in the superpotential. In the Starobinsky inflation model, the horizon exit happens in a relatively flat region of the scalar potential. However, if we introduce an USR regime near the inflection point, it will shift the horizon exit towards the minimum of the potential, thus reducing the value of $n_s$. This can be counteracted by introducing a term in the scalar potential which grows with $\varphi$, and flattens the potential in the region where the horizon exit happens. When looking at the scalar potential \eqref{V_setup} originating from the subleading terms in the superpotential, we find that the $\phi^{-1}$-term is suitable for this purpose because it is proportional to $e^{\sqrt{2/3}\varphi}$. For example, this term can be turned on by using a negative $a_0$ parameter that must be tuned in order to keep the potential flat near the  horizon exit. To summarize, we get the superpotential and the scalar potential (for the $a_4$ model) with the inclusion of $a_0$ as follows:
\begin{gather}
W=A[a_0+T^2+a_3T^3+a_4T^4+S(T+b_2T^2)]~,\label{W_a4_a0}\\
\begin{aligned}
V/A^2 &=-a_0\phi^{-1}+\tfrac{1}{2}\left(\tfrac{1}{6}-3a_0a_3\right)-\tfrac{1}{6}(2+12a_0a_4-b_2)\phi\\
&\hspace{1cm}-\tfrac{1}{12}(6a_3-b_2^2)\phi^2-\tfrac{1}{3}a_4\phi^3+\tfrac{1}{2}a_3a_4\phi^4+\tfrac{2}{3}a_4^2\phi^5~.
\end{aligned}\label{V_a4_a0}
\end{gather}
The value of $a_0$ can be chosen to raise $n_s$, the $a_3$ controls the shape of the potential near the inflection point, the $a_4$ is fixed by the Minkowski vacuum equations, and the $b_2$ controls the height of the inflection point. As for the power spectrum enhancement in the single-field near-inflection-point models, it depends on both the height of the inflection point, and the shape of the potential near it. Therefore, it is a combination of $a_3$ and $b_2$ that controls the power spectrum peak, which we also confirm numerically. The power spectrum of scalar perturbations is derived  in the next Section.

\begin{figure}
\centering
  \includegraphics[width=.45\linewidth]{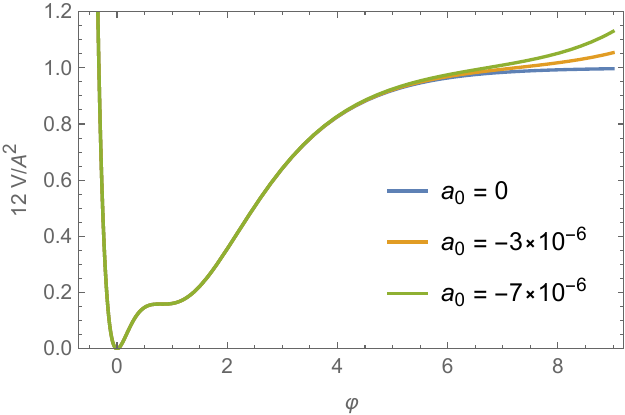}
\captionsetup{width=1\linewidth}
\caption{The scalar potential \eqref{V_a4_a0} after turning on $a_0$. The $b_2=-1/2$, the $a_3$ and $a_4$ are given in Table \ref{Tab_a4_a0}. The CMB scales leave the horizon when $\phi\approx 6$, as can be seen from Fig.~\ref{Fig_a4_sol}. The evolution of $\varphi$, $H$ and SR parameters is nearly the same for all three choices of $a_0$.}
\label{Fig_V_a4_a0}
\end{figure}

Figure \ref{Fig_V_a4_a0} shows how small negative values of $a_0$ change the scalar potential at large $\varphi$ defined by
$\phi=\langle\phi\rangle e^{-\sqrt{2/3}\varphi}$. The parameters $a_3$ and $a_4$ are fixed in Table \ref{Tab_a4_a0} by demanding $\Delta N_2=20$ and a Minkowski vacuum. Small changes between the three sets of parameters are due to the small variations in $a_0$. The evolution of $\varphi$, $H$, $\epsilon$, and $\eta$ in the three cases is nearly the same as in Fig.~\ref{Fig_a4_sol}.~\footnote{Though adding a small value to $a_0$ modifies the Starobinsky-like inflationary plateau for large $\varphi$, we find no noticeable increase in the dependence of the inflationary solutions upon initial conditions.} Table \ref{Tab_a4_a0}  shows the values of $n_s$ and $r$ for the given parameter sets. When $a_0=-7\times 10^{-6}$, the spectral tilt $n_s$ is already within $1\sigma$ CMB limits, which is the significant improvement compared to $a_0=0$ case. As for the tensor-to-scalar ratio $r$, it tends to larger values with increasing $a_0$, though still within the current CMB bounds \cite{BICEP:2021xfz,Tristram:2021tvh}. Here we  used the duration $\Delta N_2=20$ of the USR as an example. In the next Section we give the specific values of the parameters in order to obtain PBH with the masses of $10^{18}-10^{22}$ g, suitable for the whole DM, where slightly larger values of $\Delta N_2$ are favored in both ($a_4$ and $b_3$) models.

We find that the $b_3$ model of Eq.~\eqref{W_IPM_b3} leads to nearly the same results for the inflationary dynamics and observables, like the $a_4$ model. In particular, adding a small negative $a_0$-parameter to the $b_3$ model helps to raise the value of $n_s$ when $\Delta N_2$ is around $20$.

\begin{table}[hbt!]
\centering
\begin{tabular}{r r r r r r}
\toprule
$a_0$ & $a_3$ & $a_4$ & $b_2$ & $n_s$ & $r$ \\
\hline
$0$ & $-1.5157647$ & $0.774006
$ & $-1/2$ & $0.9430$ & $0.0093$ \\
$-3\times 10^{-6}$ & $-1.5157049$ & $0.773923
$ & $-1/2$ & $0.9507$ & $0.0107$ \\
$-7\times 10^{-6}$ & $-1.5156251$ & $0.773813
$ & $-1/2$ & $0.9614$ & $0.0129$ \\
\bottomrule
\end{tabular}
\captionsetup{width=1\linewidth}
\caption{The parameter sets for three different values of $a_0$, showing the impact of a small negative $a_0$ on the spectral tilt $n_s$ and the tensor-to-scalar ratio $r$. The $a_3$ and $a_4$ are tuned to obtain the USR duration $\Delta N_2=20$ in all three cases.}
\label{Tab_a4_a0}
\end{table}

\section{PBH and DM in our models}

Assuming  PBH formation during the radiation era, we use the Press-Schechter formalism \cite{Press:1973iz} to estimate the PBH mass function from a given power spectrum. We use the following expressions for the PBH mass, the PBH production rate, and the density contrast \cite{Inomata:2017okj,Inomata:2017vxo}:
\begin{gather}
    M_{\rm PBH}(k)\simeq 10^{20}\left(\frac{7\times 10^{12}}{k~{\rm Mpc}}\right)^2{\rm g}~, \quad \beta_f(k)\simeq\frac{\sigma(k)}{\sqrt{2\pi}\delta_c}
    e^{-\frac{\delta^2_c}{2\sigma^2(k)}}~,\\
    \sigma^2(k)=\frac{16}{81}\int\frac{dq}{q}\left(\frac{q}{k}\right)^4e^{-q^2/k^2}P_\car(q)~,\label{PBH_production}
\end{gather}
where $\delta_c$ is the density threshold (critical density) for PBH formation. It was estimated as $\delta_c\simeq 1/3$ \cite{Carr:1975qj}, but numerical analysis gives larger values $0.41\lesssim\delta_c\lesssim 2/3$ \cite{Musco:2018rwt}. In our calculations we choose the reference value $\delta_c=0.45$. Then the PBH fraction can be estimated as
\begin{eqnarray}
    \frac{\Omega_{\rm PBH}(k)}{\Omega_{\rm DM}}\equiv f(k)\simeq\frac{1.2\times 10^{24}\beta_f(k)}{\sqrt{M_{\rm PBH}(k){\rm g}^{-1}}}~~.\label{f_PBH}
\end{eqnarray}
The numerical factor $1.2$ was obtained by assuming the Minimal Supersymmetric Standard Model (MSSM) physical degrees of freedom (for the Standard Model it becomes $1.4$). The total PBH-to-DM fraction reads
\begin{eqnarray}
    f_{\rm tot}=\int d(\log M_{\rm PBH})f(M_{\rm PBH})~.
\end{eqnarray}
In order to derive the PBH mass function from the equations above, we have to get the power spectrum of scalar perturbations. We do this numerically by solving the Mukhanov-Sasaki (MS) equation \cite{Mukhanov:1985rz,Sasaki:1986hm} given in Appendix.

\begin{figure}
\centering
  \includegraphics[width=.7\linewidth]{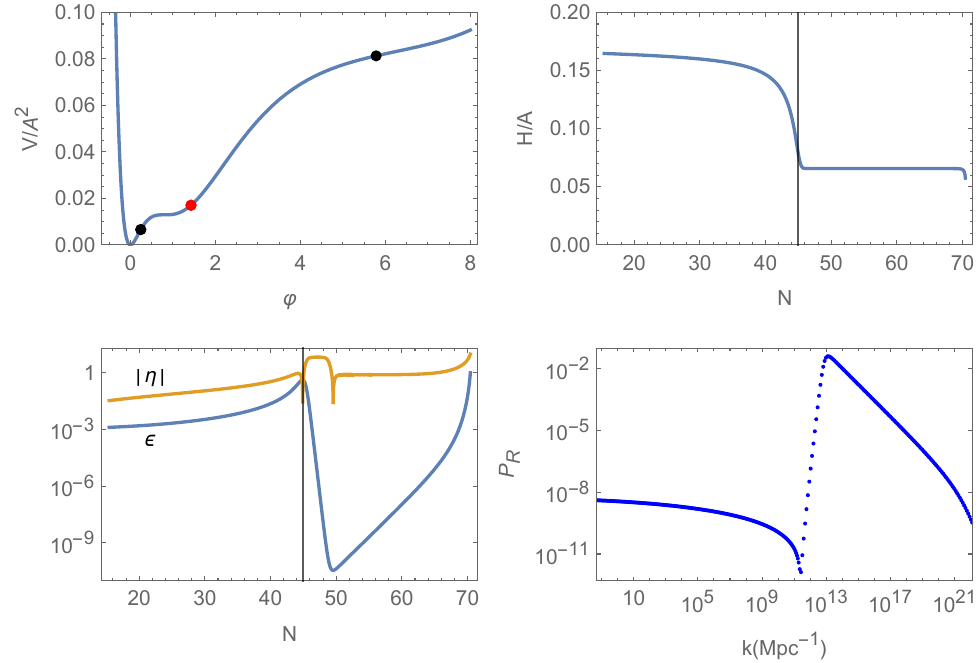}
\captionsetup{width=1\linewidth}
\caption{The inflationary solution in the $a_4$ model with the parameter choice \eqref{a4_par_PBH}. The top-left plot is the scalar potential with the black dots representing the start and the end of the last $55$ e-folds, and the red dot showing the end of the first (slow-roll) stage. The top-right and bottom-left plots show the Hubble function and the slow-roll parameters, respectively (the vertical lines show the end of the first stage). The bottom-right plot is the power spectrum of scalar perturbations, where the large peak is near the scale $10^{13}$ Mpc$^{-1}$.}
\label{Fig_PBH_sol}
\end{figure}

\begin{figure}
\centering
  \includegraphics[width=.7\linewidth]{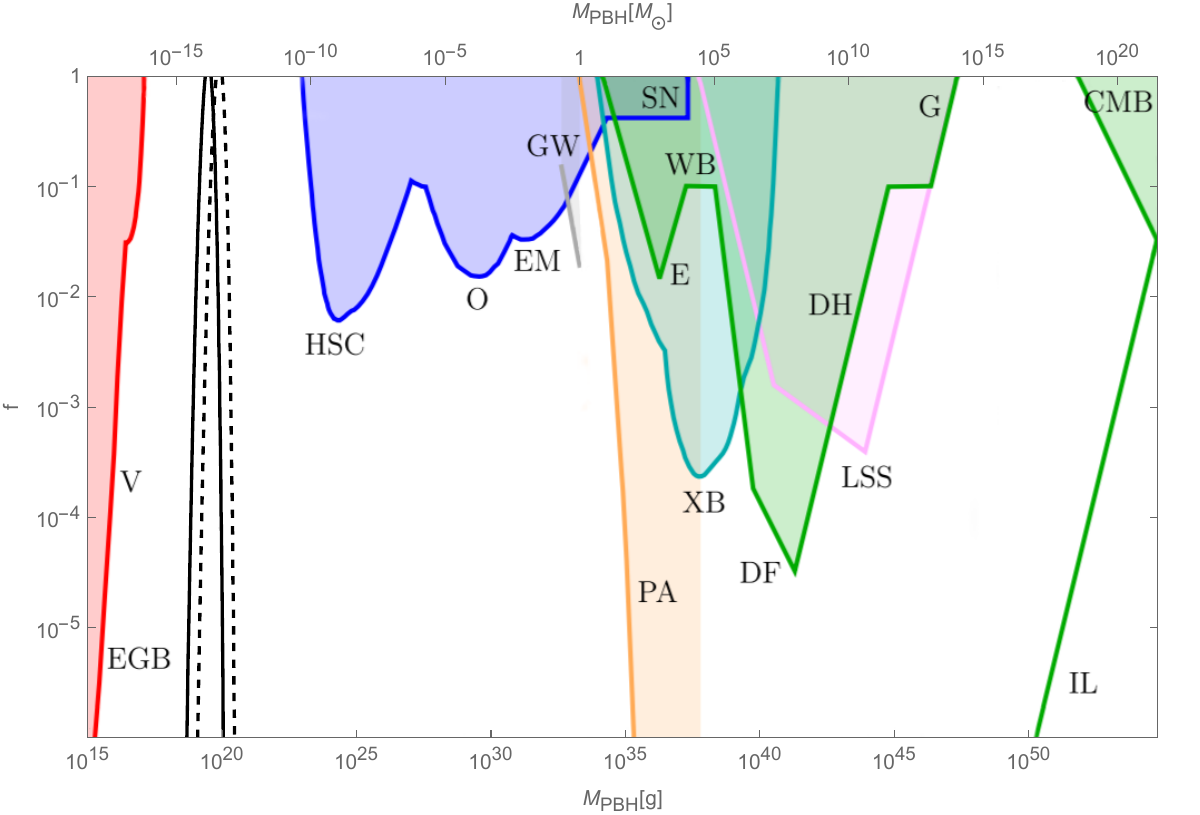}
\captionsetup{width=1\linewidth}
\caption{The PBH-to-DM mass function in the $a_4$ model with the parameters \eqref{a4_par_PBH} (the solid black curve), and in the $b_3$ model with the parameters \eqref{b3_par_PBH} (the dashed curve). In both cases, $f_{\rm tot}=1$. The background  of observational constraints is taken from Refs.~\cite{Carr:2020gox,Escriva:2022duf}.}
\label{Fig_f_PBH}
\end{figure}

First, we consider the $a_4$ model defined by the superpotential \eqref{W_a4_a0} including the $a_0$ parameter, and find the parameter choice giving rise to $f_{\rm tot}=1$, i.e. the PBH as the whole DM, while keeping the acceptable values of $n_s$ and $r$,~\footnote{When $f_{\rm tot}=1$, this parameter choice is just an example that is not unique.}
\begin{equation}\label{a4_par_PBH}
a_0=-1.4\times 10^{-5}~,~~~a_3=-1.487732305~,~~~a_4=0.75173~, ~~~b_2=-0.48~.
\end{equation}
This example leads to the duration of the USR stage $\Delta N_2=25.53$, and the inflationary observables
\begin{equation}
n_s=0.9636~,~~~r=0.0208~.
\end{equation}
The corresponding numerical plots are shown in Fig.~\ref{Fig_PBH_sol}, including the scalar potential (top-left), the Hubble function (top-right), the slow-roll parameters (bottom-left), and the power spectrum (bottom-right). The latter shows a large enhancement (slightly exceeding $P_{\car}=10^{-2}$) in the power spectrum near the scale $k=10^{13}$ Mpc$^{-1}$. We use this power spectrum in Eq.~\eqref{PBH_production} to calculate the density contrast and eventually the PBH mass function \eqref{f_PBH}, which is shown in Fig.~\ref{Fig_f_PBH} as the solid black curve. The PBH fraction peaks near $10^{19}$ g.

Similar results with the PBH-DM scenario are obtained from the $b_3$ model as well. We extend the model  in 
Eq.~\eqref{W_IPM_b3} by adding the $a_0$-term as
\begin{gather}
W=A[a_0+T^2+a_3T^3+S(T+b_2T^2+b_3T^3)]~,\label{W_b3_a0}\\
\begin{aligned}
V/A^2=& -a_0\phi^{-1}+\tfrac{1}{2}\left(\tfrac{1}{6}-3a_0a_3\right)-\tfrac{1}{6}(2-b_2)\phi\\
&-\tfrac{1}{12}(6a_3-b_2^2-2b_3)\phi^2+\tfrac{1}{6}b_2b_3\phi^3+\tfrac{1}{12}b_3^2\phi^4~.
\end{aligned}\label{V_b3_a0}
\end{gather}

To realize the whole PBH-DM scenario, we take the following parameters:
\begin{equation}\label{b3_par_PBH}
a_0=-3.7\times 10^{-5}~,~~~a_3=-0.1566928828~,~~~b_2=1.044~,~~~ b_3=-0.249931~,
\end{equation}
which lead to the inflationary (CMB) predictions
\begin{equation}
n_s=0.9616~,~~~r=0.0216~,
\end{equation}
and the PBH mass function given by the dashed curve in Fig.~\ref{Fig_f_PBH}, where the peak is located near  $10^{20}$ g. The USR stage lasts for $\Delta N_2=25.94$ e-folds. We do not show plots of the inflationary solution in this case because they are nearly identical to Fig.~\ref{Fig_PBH_sol}, including the shape of the power spectrum.

\begin{figure}
\centering
  \includegraphics[width=.5\linewidth]{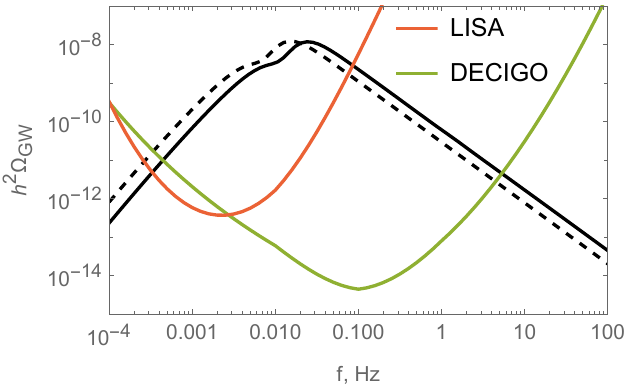}
\captionsetup{width=1\linewidth}
\caption{The GW background predicted by the model $a_4$ with the parameter choice \eqref{a4_par_PBH} (the solid black curve), and by the model $b_3$ with the parameter choice \eqref{b3_par_PBH} (the dashed curve).}
\label{Fig_GW}
\end{figure}

The low-mass PBH-DM  formation is  known to produce the notable stochastic GW background induced by  enhanced scalar perturbations, within the frequencies of the planned space-based detectors such as LISA \cite{LISA} and DECIGO \cite{DEC}. To calculate this GW background, we use the formalism of 
Refs.~\cite{Espinosa:2018eve,Kohri:2018awv,Bartolo:2018evs}, see also Section 7 of Ref.~\cite{Aldabergenov:2022rfc}. The resulting GW density is shown in Fig.~\ref{Fig_GW}. The induced primordial GW signal caused by scalaron would be a clear signature of PBH, being complementary to another primordial GW signal induced by PBH Poisson fluctuations in Starobinsky gravity \cite{Papanikolaou:2021uhe} and VAS supergravity, see Ref.~\cite{Kawai:2022emp} also.

The canonical inflaton and axion masses, and the SUSY breaking parameters in our two examples are given in Table \ref{Tab_masses}. There is a small difference in the mass values, while the SUSY breaking scale is slightly higher in the $b_3$ model. The axion mass is higher than the inflaton mass in both cases,~\footnote{We find that the effective axion mass at the horizon exit is $6.92\times 10^{13}$ GeV and $7.04\times 10^{13}$ GeV in the models $a_4$ and $b_3$, respectively, while it does not significantly change during the whole inflation. This means that the axion is stabilized throughout the inflationary history. It is to be compared to the Hubble function value at the horizon exit, which is $H_*=3.49\times 10^{13}$ GeV and $H_*=3.55\times 10^{13}$ GeV in the models $a_4$ and $b_3$, respectively.} while inflaton can be lighter or heavier than gravitino depending upon the model. Since a precise location of the peak varies within the available window in Fig.~\ref{Fig_f_PBH}, the masses in Table \ref{Tab_masses} slightly vary as well.

\begin{table}[hbt!]
\centering
\begin{tabular}{l | r r r r r}
\toprule
 & $m_\varphi/{\rm GeV}$ & $m_\tau/{\rm GeV}$ & $\langle m_{3/2}\rangle/{\rm GeV}$ & $\langle|F^S|\rangle/{\rm GeV}^2$ & $\langle|F^T|\rangle/{\rm GeV}^2$ \\
\hline
Model $a_4$ & $1.07\times 10^{14}$ & $1.22\times 10^{14}$ & $1.97\times 10^{13}$ & $7.82\times 10^{31}$ & $1.11\times 10^{32}$ \\
Model $b_3$ & $5.05\times 10^{13}$ & $3.30\times 10^{14}$ & $6.87\times 10^{13}$ & $2.75\times 10^{32}$ & $1.99\times 10^{32}$ \\
\bottomrule
\end{tabular}
\captionsetup{width=1\linewidth}
\caption{The canonical masses of inflaton $\varphi$ and axion (sinflaton) $\tau$, the gravitino mass, and the SUSY breaking F-field VEV in our two PBH-as-whole-DM models described in this Section.}
\label{Tab_masses}
\end{table}

\section{Conclusion}

In this work we studied PBH production from (effectively) single-field ultra-slow-roll phase in the framework of Volkov--Akulov--Starobinsky supergravity, which combines non-linearly realized spontaneously broken $N=1$ supersymmetry and Starobinsky inflation. The VAS supergravity is based on the no-scale K\"ahler potential \eqref{K_no-scale} including the inflaton chiral superfield $T$ and the nilpotent superfield $S$, and the bilinear superpotential \cite{Antoniadis:2014oya,Aldabergenov:2020atd}. In order to introduce a near-inflection point to the scalar potential for ultra-slow-roll phase, we generalize the superpotential to a general polynomial of the form \eqref{W_gen} without changing the K\"ahler potential. We find that the superpotential must include at least a quartic term, either $W\supset T^4$ or $W\supset ST^3$, in order to support an inflection point. We give the specific examples for each quartic term, where PBH describe whole dark matter.

Our approach offers several advantages over the other supergravity-based Starobinsky-like PBH-DM models known in the literature 
\cite{Dalianis:2014aya,Aldabergenov:2020bpt,Nanopoulos:2020nnh,Kawai:2022emp}.
First, spontaneous SUSY breaking is automatically included by imposing the nilpotency constraint on $S$. It makes the effective 
low-energy field theory blind to the ultra-violet (UV) dynamics that gives rise to the nilpotency constraint. Second, flexibility of our model (given by its superpotential) allows us to keep the inflationary observables $n_s$ and $r$ within the current CMB bounds (at the $1\sigma$ confidence level), while producing the whole PBH-DM in the observationally allowed asteroid-mass window. Our models  predict potentially observable gravitational waves (GW) from two different origins: (i) primordial GW leading to a relatively large tensor-to-scalar ratio $r$, which could be tested by more precise CMB measurements such as 
LiteBIRD project \cite{LiteBIRD:2022cnt} in the future, and (ii) large scalar-induced GW that could be tested by the space-based GW interferometers  such as LISA \cite{LISA} and DECIGO \cite{DEC}.

Another relevant phenomenological aspect of our models is spontaneous  SUSY breaking whose scale is directly related to the scale of inflation, around $ 10^{13}-10^{14}$ GeV.

\section*{Acknowledgements}

YA was supported by Thailand NSRF via the Program Management Unit for Human Resources and Institutional Development, Research and Innovation, under the grant Nos. B01F650006 and B05F650021. SVK was supported by the World Premier International Research Center Initiative (WPI Initiative), MEXT, Japan, the Japanese Society for Promotion of Science under the grant No.~22K03624, and the Tomsk Polytechnic University Development Program Priority-2030-NIP/EB-004-0000-2022. 

The authors thank Sayantan Choudhury, Guillem Domenech, Kazunori Kohri, Jinsu Kim, Florian Kuehnel, Ahmad Moursy, and Theodoros Papanikolaou for discussions and correspondence.

\section*{Appendix: Mukhanov-Sasaki equation}
\addcontentsline{toc}{section}{\protect\numberline{}Appendix: Mukhanov-Sasaki equation}

The Mukhanov-Sasaki (MS) equation \cite{Mukhanov:1985rz,Sasaki:1986hm} describes the evolution of scalar perturbations,
\begin{equation}
\left(\frac{d^2}{d\tau^2}+k^2-\frac{d^2z}{zd\tau^2}\right)u_k=0~,
\end{equation}
where $\tau$ is conformal time ($d\tau=dt/a$), $z\equiv\frac{d\phi}{Hd\tau}$, and $u\equiv z\car$ ($u_k$ is its $k$-mode), for a comoving curvature perturbation $\car$.

It is convenient to rewrite the MS equation in terms of the e-folds variable $N$, see e.g., Refs. \cite{Ballesteros:2017fsr,Ketov:2022zhp}. Here we use the MS equation in the form
\begin{equation}
u''_k+(1-\epsilon)u_k+\left[\frac{k^2}{(aH)^2}+(1+\tfrac{1}{2}\eta)(\epsilon-\tfrac{1}{2}\eta-2)-\tfrac{1}{2}\eta'\right]u_k=0~,
\end{equation}
with the definitions of the slow-roll parameters as $\epsilon\equiv -H'/H$ and $\eta\equiv \epsilon'/\epsilon$. The $u_k(N)$ begins to evolve deep inside the horizon, when $k\gg aH$, with the Bunch-Davies initial condition
\begin{equation}
u_k=\frac{e^{-ik\tau}}{\sqrt{2k}}~,
\end{equation}
where conformal time is related to $N$ as $d\tau=dN/(aH)$. The solution $u_k(N)$ for each mode $k$ is used to build the power spectrum of scalar perturbations,
\begin{equation}
P_{\car}=\frac{k^3}{2\pi^2}\left|\frac{u_k}{z}\right|^2_{k\ll aH}~,
\end{equation}
to be evaluated at a later time when $k\ll aH$, in order to allow the mode $u_k$ to stabilize at a constant value.

\providecommand{\href}[2]{#2}\begingroup\raggedright\endgroup

\end{document}